\documentclass[12pt]{article}
\usepackage{epsfig, amssymb}
\usepackage{graphicx,epsfig} 
\begin{document}
\begin{titlepage}
\thispagestyle{empty}
\begin{flushright}
\end{flushright}

\bigskip

\begin{center}
\noindent{\Large \textbf
{Running Shear Viscosities in Anisotropic Holographic Superfluids }}\\
\vspace{2cm} \noindent{
Jae-Hyuk Oh\footnote{e-mail:jack.jaehyuk.oh@gmail.com}}

\vspace{1cm}
  {\it
 Harish-Chandra Research Institute \\
 
 Chhatnag Road, Jhunsi, Allahabad-211019, India\\
 }
 
%
\end{center}

\vspace{0.3cm}
\begin{abstract}
We have examined holographic renormalization group($RG$) flows of the shear viscosities in anisotropic holographic 
superfluids via their gravity duals, Einstein-$SU(2)$ Yang-Mills system. 
In anisotropic phase, below the critical temperature $T_c$, the $SO(3)$ isometry(spatial rotation) in the dual gravity system 
is broken down to the residual $SO(2)$. 
The shear viscosities in the symmetry broken directions of the conformal fluids defined on $AdS$ boundary present non-universal values which depend on
the chemical potential $\mu$ and temperature $T$ of the system and also 
satisfy non-trivial holographic $RG$-flow equations. 
The shear viscosities flow down to the specific values in $IR$ 
region, in fact which are given  
by the ratios of the metric components in the symmetry unbroken direction to those 
in the broken directions, evaluated at the black brane horizon in the dual gravity system.


\end{abstract}
\end{titlepage}
\newpage

\section{Introduction}
The Gauge/gravity duality has shed light on strongly coupled field theories. 
Especially, Fluid/gravity duality is widely studied and it provides much useful information about conformal fluid dynamics in
the effectively long wavelength limit. Many crucial quantities in real time formalism in conformal fluids 
can have been obtained via retarded(advanced) Green's function(s) from their gravity duals \cite{Son1,Son2}.

The most celebrated example from holographic fluid dynamics is known as the ratio of the shear viscosity $\eta$ to entropy density
$s$. The ratio seems to be universal for many conformal fluids, $\frac{\eta}{s}=\frac{1}{4\pi}$ \cite{Son3,Alex1,Alex2,Liu1}, 
of which holographic duals are
Einstein gravities in asymptotically $AdS$ space. So far, the only violation of this universality appears when string effects or quantum effects are 
taken into account in the dual gravity system\cite{Alex3,Alex4,Aninda1}. 

One interesting direction to investigate the shear viscosities in boundary conformal fluids is 
studying their holographic renormalization group($RG$) flow equations. It has been argued that Wilsonian $RG$ flow is 
consistent with holographic $RG$ in \cite{Polchinski1,Hong11}. 
In the dual gravity system, the radial direction, $r$ of $AdS$ space
is identified with Wilsonian $RG$-direction and that radial direction is related to the energy scale of the dual fluids. 
$AdS$ boundary is treated as $UV$-region whereas the black brane horizon is as $IR$-region. 
In between, one can define a fluid dynamics at intermediate energy scale and such conformal fluids are defined on a hyper surface located 
at $r= \tilde r$, where $ r_h <\tilde r < \infty$, $r_h$ is the black brane horizon and $AdS$ boundary is located at $r=\infty$.

In principle, holographic $RG$ flow  and the radial evolution of bulk Einstein equations of motion are different.
In \cite{Polchinski1,Hong11,Zhou2}, however,  the authors have provided a general 
proof that two flows are indeed equivalent in the limit of classical gravities.
Therefore, once one solves the bulk equations of motion, 
the corresponding flows are completely known.

For the cases so far\cite{Liu1,Zhou1}, the shear viscosity does not run along 
the radial direction.  
From explicit calculation, it is obtained that $\partial_{\tilde r} \eta=0$ for
these cases.
In \cite{Strominger1}, the shear viscosity runs, but the entropy density also runs in such a way 
that the shear viscosity to the 
entropy density ratio does not change.
In \cite{Liu1}, the authors argue that the physical reason why the shear viscosity shows trivial flow may be found 
in $membrane\ paradigm$. The $membrane\ paradigm$ may provide an argument that the linear response of the boundary fluid dynamics 
is completely captured by that of the horizon fluid dynamics in the small frequency limit. 
The transport coefficients on the boundary fluids can be expressed in terms of those on the horizon fluids only. 
The shear viscosity is expected to be an apparent example of this.

However, we will provide an example of non-trivial $RG$ flows 
of the shear viscosities from anisotropic conformal fluid dynamics which has been suggested recently
\cite{Basu:2009vv,Johanna1}.
This fluid system displays an order parameter which depends on a certain spatial direction below its critical temperature 
$T_c$  and becomes $p$ wave superfluid, 
in which the ratio in that direction does not show the universal value
\cite{Johanna2,Natsuume:2010ky}
\footnote{The universality of the shear viscosity and the entropy density ratio
is violated for some other cases too, as higher derivative gravity theories \cite{Myers1,Paulos1,Myers2,Kats1,Sun1,Sera1,Sera3},
Einstein-Axion-dilaton system\cite{Anton1} and a study on non-trivial temperature dependence of the ratio\cite{Sera2}}. 
The gravity dual of this conformal fluids is
Einstein-$SU(2)$ Yang-Mills system
\footnote{See \cite{Manvelyan:2008sv,Gubser:2008zu,Gubser:2008wv,Roberts:2008ns} for pioneering works 
on connection between Einstein-$SU(2)$-Yang-Mills and $p$-wave holographic superfluids.}. 
It is an exact solution of the bulk action with an $AdS$ black brane in 5-dimension with non-zero chemical 
potential turned on for the temporal component of the gauge potential proportional to $\sigma_3$ of $SU(2)$ 
gauge group. The boundary metric enjoys 
$SO(3)$ global symmetry (as the spatial rotation in $x,y$ and $z$ directions in our coordinate system). At high temperature (equivalently small chemical potential $\mu$), 
the system stays in the
isometric phase. However at a certain chemical potential $\mu=\mu_c$(equivalently, a certain temperature $T=T_c$), 
$SO(3)$ symmetry is spontaneously broken into residual $SO(2)$(rotations in $y$ and $z$ directions) because
one of the spatial components($x$ directional component) of the Yang-Mills fields develops a non-trivial zero mode (as a
solution of the linearized Yang-Mills field equations) and that mode is
thermodynamically more favorable than the trivial zero mode(isotropic phase). 
It turns out that in the region of $\mu>\mu_c$, this mode condenses and there is a new anisotropic superfluid phase.


In this note, we have obtained flow equations of the shear viscosities in symmetry broken direction in the anisotropic phase 
using the radial flows of the bulk equations of motion. 
It turns out that the flow equations are not trivial, since they contain contributions from 
interactions with perturbative Yang-Mills fields. 


The shear viscosities will non-trivially flow into some specific values deeply in $IR$ region, which would correspond 
to the black brane horizon in the dual gravity system.  It turns out that the shear viscosities at the black brane horizon are expressed in terms of 
the ratios of the metric components in the symmetry unbroken directions to the those 
in the broken directions. For our case, 
the metric of the bulk spacetime is still diagonal in anisotropic phase. Therefore, for example, the shear viscosity $\eta_{xy}$ will be given by
\begin{equation}
\label{SHEAR IN HORIZON}
\eta_{xy}=\frac{1}{2\kappa^2_5}\left.\frac{G_{yy}}{G_{xx}}\right|_{\rm at\ the\ horizon},
\end{equation}
where $\kappa_5$ is the 5-dimensional gravity constant, $G_{xx}$ and $G_{yy}$ are diagonal metric components of the bulk spacetime 
and $\frac{1}{2\kappa^2_5}$ is the universal value of the shear viscosity.

The reason for this behaviors in $IR$ region is due to the causal boundary condition 
at the black brane horizon. 
Near horizon, the
fast oscillating factor, $(r-r_h)^{-i\nu\beta}$ of the gravitational perturbations which contribute to the shear viscosity 
calculations has dominant contribution and 
the other regular factors are relatively suppressed, where $\beta$ is a positive real number and
$\nu$ is frequency of the fields. 
In the equation of motion of 
the gravitational perturbations, the interaction terms from Yang-Mills fields 
are relatively suppressed near horizon and 
the equation becomes the same form of the near horizon limit of that in isotropic phase. 
Since $\beta$ depends only on the background metric evaluated at the horizon
\footnote{In fact, $\beta$ is proportional to the inverse of the Hawking temperature $T$.}, the shear viscosities will do as such. 
In some literatures (e.g see\cite{Anton1}), the form of the shear 
viscosity as Eq(\ref{SHEAR IN HORIZON}) at the horizon is expected but they are the cases for trivial $RG$ flows. 
We have shown that this is true for the case of the non-trivial flows too.


In the last section in this note, we will provide a specific example for this argument. Using the analytic solutions in 
Einstein-$SU(2)$ Yang-Mills obtained in \cite{Oh1}, we explicitly show that Eq(\ref{SHEAR IN HORIZON}) is  valid near critical point.
This analytic solution can be obtained only when the anisotropic order parameter is small and Yang-Mills coupling is very large.
However, the argument is quite general.
Therefore, for any temperature, 
we expect that the shear viscosity behaves as Eq(\ref{SHEAR IN HORIZON}) in $IR$ region.

This note is organized as follows. In Sec.\ref{Analytic Solutions in Large Coupling Expansion and Non-universal Values of the Shear viscosity},
we briefly review Einstein-$SU(2)$ Yang-Mills system and its analytic solutions. This analytic solutions will be used to
provide an example for our claim in Sec.\ref{General}.
In Sec.\ref{RG Flow of the Shear Viscosity in Symmetry Broken Direction}, 
we will derive $RG$ equation for the shear viscosities in the symmetry broken direction and
explore their properties.

\section{Analytic Solutions in Einstein-$SU(2)$ Yang-Mills System and Non-universal Values of the Shear Viscosity}
\label{Analytic Solutions in Large Coupling Expansion and Non-universal Values of the Shear viscosity}
In this section, we briefly review the analytic solutions in Einstein-$SU(2)$-Yang-Mills, 
the dual gravity system of anisotropic superfluids. We mostly follow \cite{Oh1} for the discussion and this solution will be used to
provide a concrete example of non-trivial $RG$ flows of the shear viscosities in this system.
\subsection{Holographic Setup and Analytic Solutions in Large Coupling Expansion}
\label{Holographic Setup and Analytic Solutions in Large Coupling Expansion}
The authors in \cite{Oh1} consider Einstein-$SU(2)$ Yang-Mills system of which space-time is asymptotically $AdS_{5}$. The action is
\begin{equation}
S=\int d^{5}x \sqrt{-G} \left(  \frac{1}{\kappa^2_5}(R+\frac{12}{L^2}) - \frac{1}{4g^2}F^{a}_{MN}F^{aMN} \right),
\end{equation}
where $M$, $N$... are 5-dimensional space-time indices, $a$.. are $SU(2)$ indices and $g$ is Yang-Mills coupling. 
For further discussion, we choose $L=1$. Yang-Mills field strength $F^{a}_{MN}$ is given by
\begin{equation}
F^{a}_{MN}=\partial_{M}A^{a}_{N}-\partial_{N}A^{a}_{M}-\epsilon^{abc}A^{b}_{M}A^{c}_{N},
\end{equation}
where $\epsilon^{abc}$ is anti-symmetric tensor with $\epsilon^{123}=1$.
The equations of motion from the action are obtained as
\begin{eqnarray}
\label{W-equation}
W_{MN}&\equiv&R_{MN}+4G_{MN}-\kappa^{2}_{5}\left( T_{MN}-\frac{1}{3} T^{P}_{P}G_{MN} \right)=0, \\ 
\label{Y-equation}
Y^{aN}&\equiv&\nabla_{M}F^{aMN}-\epsilon^{abc}A^{b}_{M}F^{cMN}=0,
\end{eqnarray}
where $T_{MN}$ is the energy-momentum tensor, of which form is
\begin{equation}
T_{MN}=\frac{1}{g^2}\left( F_{MP}^{a}F_{N}^{Pa} -\frac{1}{4}F_{PQa}F^{PQa}G_{MN} \right).
\end{equation}
The ansatz for the metric and Yang-Mills field are given by
\begin{eqnarray}
\label{gauge and metric ansatz}
A&=&\phi(r)\tau^{3}dt+\omega(r)\tau^{1}dx, \\ \nonumber
ds^2&=&-N(r)\sigma^2(r)dt^2+\frac{dr^2}{N(r)}+r^2f^{-4}(r)dx^2+r^2f^{2}(r)\left( dy^2 + dz^2 \right),
\end{eqnarray}
where $\tau^{a}=\frac{s^{a}}{2}$ and $s^{a}$ are Pauli-matrices. We never discuss the detailed forms of equations of motion here but just provide their solutions
\footnote{For the precise equations of motion, see Sec.2 in \cite{Oh1}}.

A known exact solution of the equations of motion is asymptotically $AdS_5$, charged-black-brane solution, of which forms are
\begin{eqnarray}
\label{the zeroth order in varepsilon}
\phi(r)&=&\tilde{\mu} (1-\frac{r^{2}_{h}}{r^2}), {\ \ }\omega(r)=0, \\ \nonumber
\sigma(r)&=&f(r)=1 {\rm \ \ and\ }N(r) = N_{0}(r) \equiv  r^2 -\frac{m}{r^2}+\frac{2\tilde{\mu}^2\alpha^2r^4_{h}}{3r^4} ,
\end{eqnarray}
where $\tilde{\mu}$ is chemical potential, $r_h$ is the black brane horizon, $\alpha^2=\frac{\kappa^2_5}{g^2}$ and $m=r^4_{h}+\frac{2\mu^2\alpha^2r^2_{h}}{3}$. 
In the infinite Yang-Mills coupling limit as $g \rightarrow \infty$, the last term in $N(r)$ vanishes and the solution becomes uncharged. 

To explore this system near critical point, $\tilde \mu=\tilde \mu_c=4r_h$, by analytic method
\footnote{For numerical approaches, see \cite{Johanna1,Johanna2}}
, the authors in \cite{Oh1}  develop double expansion to the metric fields and Yang-Mills field  order by order in $\varepsilon \tilde{D}_{1}$ 
and $\alpha^2$. 
$\varepsilon$ is dimensionless small parameter and $\tilde{D}_{1}$ is the $SO(3)$ rotational symmetry breaking order parameter, appearing in the non-trivial zero mode of $\omega(r)$,
\begin{equation}
\omega(r)=\varepsilon\frac{\tilde{D}_{1}r^2}{(r^2 +1)^2}+O(\varepsilon^2 \tilde D^2_1).
\end{equation}
It is convenient to choose a convention that the horizon of the black brane is located at $r=1$ 
by scaling that $r \rightarrow r_{h} r$, $\{t,x,y,z \} \rightarrow \frac{1}{r_{h}}\{t,x,y,z \}$ 
and defining a new chemical potential $\mu \equiv \frac{\tilde{\mu}}{r_{h}}$.
The equations of motion enjoy certain scaling symmetry\cite{Johanna1,Basu:2009vv}. 
By these, we can choose the asymptotic values of $\sigma(r=\infty)=1$ and $f(r=\infty)=1$ 
on the large $r$ boundary for the space-time to become asymptotically $AdS_{5}$. 
The value of chemical potential is taken to be $\mu=4$ for the dual boundary field theory system 
to be at the critical point. 

To obtain the corrections in this double expansion, any appearing fields, $a(r)$ 
in the ansatz(\ref{gauge and metric ansatz}) can be expanded as
\begin{equation}
\label{varepsilon expansion}
a(r)=a_{0}(r)+\varepsilon a_{1}(r) +\varepsilon^2 a_{2}(r)...
\end{equation}
Each $a_i$ in the above expression can also be expanded as
\begin{equation}
\label{alpha expansion}
a_{i}(r)=a_{i,0}(r)+\alpha^2a_{i,2}+\alpha^4a_{i,4}(r)...
\end{equation}
The zeroth order solutions in $\varepsilon$ is given in Eq(\ref{the zeroth order in varepsilon}), 
where only $N_{0}$ contains the subleading correction of 
$\alpha^2$ in the sense of the above expansion. $N_{0,2}=\frac{32}{3}\left( \frac{1}{r^4}-\frac{1}{r^2} \right)$
and $N_{0,i}=0$ for $i=4,6...$. 
It turns out that the non-trivial leading order correction is $O(\varepsilon^2 \alpha^2)$.
The leading corrections for the metric are given by
\begin{eqnarray}
\label{background_metric_ea}
\sigma(r)&=&1-\varepsilon^2\alpha^2\frac{2\tilde{D}^2_{1}}{9(1+r^2)^3}, {\ \ }f(r)=1-\varepsilon^2\alpha^2\frac{\tilde{D}^2_{1}(1-2r^2)}{18(1+r^2)^4} \\ \nonumber
{\rm and \ \ }N(r)&=&r^2-\frac{1}{r^2} +\frac{32\alpha^2}{3}\left( \frac{1}{r^4}-\frac{1}{r^2} \right)
-\varepsilon^2\alpha^2\frac{4\tilde{D}^2_{1}}{9r^2}\left( \frac{1+2r^2}{r^2(1+r^2)^3}-\frac{3r^2}{2(1+r^2)^2} \right. \\ \nonumber
&+&\left.\frac{281}{560}\left(1-\frac{1}{r^2} \right)\right).
\end{eqnarray}
The solutions of Yang-Mills fields\footnote{Any subleading corrections of Yang-Mills field in $\alpha^2$ would not contribute to the leading back reactions to the metric. The aim of the calculation in \cite{Oh1} is to get metric corrections up to non-trivial leading order corrections, $O(\alpha^2 \varepsilon^2)$.
Therefore, it is enough that the Yang-Mills field solutions is evaluated up to $\phi_{i,0}$ and $\omega_{i,0}$ only.}
\begin{eqnarray}
\omega(r)&=&\varepsilon\frac{\tilde{D}_{1}r^2}{(r^2 +1)^2} + O(\varepsilon^2), \\ 
\phi(r)&=&4(1-\frac{1}{r^2})+\frac{\varepsilon^2\tilde{D}^2_{1}}{4}\left(\frac{(1+2r^2)}{3r^2 (1+r^2)^3}-\frac{1}{8}
+\frac{281}{1680}\left( 1-\frac{1}{r^2} \right)\right)+O(\varepsilon^3).
\end{eqnarray}

The black brane temperature is changed by the leading corrections as
\begin{equation}
 T=\frac{1}{\pi}\left( 1-\frac{16}{3}\alpha^2+\frac{17}{1260}\varepsilon^2 \alpha^2 \tilde{D}^2_{1}\right),
\end{equation}
where $T_{c} \equiv \frac{1}{\pi}\left( 1-\frac{16}{3}\alpha^2 \right)$ is the critical temperature.
The black brane entropy is given by
\begin{equation}
\label{Anisotrophic entropy}
S=\frac{2\pi}{\kappa^2_5}V_{3},
\end{equation}
where $V_{3}$ is spatial coordinate volume of the boundary space-time, $V_{3} = \int dxdydx$, in this rescaled coordinate.

\subsection{Non-Universality in Anisotropic Background}
In anisotropic background developed in the previous subsection, 
it is manifest that $SO(3)$ isometry in the background metric is broken down to $SO(2)$. 
As long as we are looking at a solution with $SO(3)$ symmetry, the universality of the ratio of entropy density 
and shear viscosity holds. This is because shear viscosities rely on the gravitational perturbations in tensor 
modes of $SO(3)$
in the dual gravity. Each tensor mode satisfies a massless scalar field equation decoupled from
one another, which can ensure the universality. 
The universality of this ratio is lost in the symmetry broken phase.
The reason is that once $SO(3)$ symmetry is broken into 
$SO(2)$, the gravitational wave modes in the symmetry
broken direction are no longer tensor modes in residual $SO(2)$. They will not be decoupled
from other fields and in fact interact with Yang-Mills fields. 
In \cite{Oh1}, the authors have computed a deviation from the universal value of the shear viscosity in the symmetry broken direction as
\begin{equation}
\label{ratio eta_yz and s}
\eta_{xy}=\eta_{xz} =\frac{1}{2\kappa^2_{5}}
\left( 1+ \frac{29}{896}\varepsilon^2 \alpha^2 \tilde{D}^2_{1}\right),
\end{equation}
using double expansion in $\varepsilon \tilde D_1$ and $\alpha^2$ together with small frequency expansion
\footnote{The shear viscosity in the direction of residual symmetry is still universal, $\eta_{yz}=\frac{1}{2\kappa^2_{5}}$ which depends on tensor modes in residual $SO(2)$.}.
This result is obtained via Kubo formula using the holographic computation of retarded Green's function 
from the perturbations, $h_{MN}$ and $\delta A^a_M$ defined as
\begin{equation}
G_{MN} =G^{(0)}_{MN}+h_{MN} {\rm \ and \ } A^{a}_M = A^{a(0)}_M + \delta A^a_M,
\end{equation}
where $G^{(0)}_{MN}$ and $A^{a(0)}_M$ are background metric and Yang-Mills fields respectively.
For example, to compute $\eta_{xy}$, one consider $h_{xy}$ together with $\delta A^1_{y}$ 
and $\delta A^2_{y}$(This is minimal set of fields interacting one another classified by residual $SO(2)$ and $Z_2$ symmetry in the symmetry broken phase). 
Their equations of motion are given by
\begin{eqnarray}
\label{Psi-equation}
0&=&\Psi_\nu^{\prime\prime}(r)+\left( \frac{1}{r} +\frac{4r}{N(r)} +\frac{6f^{\prime}(r)}{f(r)} -\frac{r\alpha^2 \phi^{\prime 2}(r)}{3N(r)\sigma^{2}(r)} \right)\Psi_\nu^{\prime}(r) +\frac{\nu^2\Psi_\nu(r)}{N^{2}(r)\sigma^{2}(r)} \\ \nonumber
&+& \frac{2\alpha^2}{r^2 f^{2}(r)}\left( \omega^{\prime}(r)\delta A_{y}^{1\prime}(r) -\frac{\omega(r)\phi^{2}(r)\delta A_{y}^{1}(r) }{N^{2}(r)\sigma^{2}(r)} +\frac{i\nu\omega(r)\phi(r)\delta A_{y}^{2}(r) }{N^{2}(r)\sigma^{2}(r)}\right), \\
\label{a1-equation}
0&=&\delta A_{y}^{1\prime\prime}(r)+\left( \frac{1}{r} -\frac{2f^{\prime}(r)}{f(r)} +\frac{N^{\prime}(r)}{N(r)} 
+ \frac{\sigma^{\prime}(r)}{\sigma(r)} \right)\delta A_{y}^{1\prime}(r)
+ \left( \frac{\nu^2+\phi^{2}(r)}{N^{2}(r)\sigma^{2}(r)}  \right)\delta A_{y}^{1}(r)\\ \nonumber
&-&f^{6}(r)\omega^{\prime}(r)\Psi_\nu^{\prime}(r)-\frac{2i\nu \phi(r)\delta A_{y}^{2}(r)}{N^{2}(r)\sigma^{2}(r)}, \\
\label{a2-equation}
0&=&\delta A_{y}^{2\prime\prime}(r)+\left( \frac{1}{r} -\frac{2f^{\prime}(r)}{f(r)} 
+\frac{N^{\prime}(r)}{N(r)} + \frac{\sigma^{\prime}(r)}{\sigma(r)} \right)\delta A_{y}^{2\prime}(r) 
+\left( \frac{\nu^{2}+\phi^{2}(r)}{N^{2}(r)\sigma^{2}(r)} \right)\delta A_{y}^{2} \\ \nonumber
&-&\frac{f^4(r)\omega^{2}(r)}{r^2 N(r)}\delta A_{y}^{2}  +\frac{i\nu\phi(r)}{N^{2}(r)\sigma^{2}(r)}(-f^{6}(r)\omega(r)\Psi_\nu(r)+2\delta A_{y}^{1}(r)),
\end{eqnarray}
where $\Psi_\nu(r)=\frac{h_{xy}}{r^2f^2(r)}$ and $\nu$ is frequency of the fields.
The equations are evaluated in frequency space and spatial momenta of the fields are turned off, $\vec{k}=0$.
The precise forms of the solutions of the weak fields are given in \cite{Oh1}
\footnote{ The solutions are quite complicated, so we would not show all these here. For detailed solutions, see Eq(34) in Sec3.2 and
Appendix.C in \cite{Oh1}. In Sec.\ref{The Exact $RG$-Flow Solution of the Shear Viscosity in the Perturbative Regim}
in this note, we need to introduce an $O(1)$ integral constant, $ \tilde A^{(1)}_{1,0}$, 
which appears in the solution of $ \delta A^1_y(r)$. This will be determined by boundary condition 
in $O(\nu^2)$, but it is
difficult to determine that analytically. So we leave this as undetermined in the following discussion. In fact, it
is not relevant for the shear viscosity computation neither at the horizon or on the $AdS$ boundary.
}.

\section{$RG$-Flows of the Shear Viscosity in Einstein-$SU(2)$ Yang-Mills System}
\label{RG Flow of the Shear Viscosity in Symmetry Broken Direction}
\subsection{$RG$ Flow of the Shear Viscosity in Symmetry Broken Direction}
\label{General}
As discussed in \cite{Liu1,Zhou1}, retarded Green's function for the gravitational perturbation, $\Psi_\nu(r)=h^y_{\ x}$ on $r=\tilde r$ 
hyper surface is given by the ratio of its canonical momentum $\Pi_\Psi(r)$ to itself, which is given by
\begin{equation}
\label{Greens function at the hyper surface}
G_{\Psi}(\tilde{r},\nu)
=\frac{\Pi_\Psi(\tilde r)}{\Psi_\nu(\tilde r)}
=\frac{\tilde r^3\sigma(\tilde r)N(\tilde r)f^6(\tilde r)\partial_{\tilde r}\Psi_{\nu}(\tilde r)}{2\kappa^2_5 \Psi_\nu(\tilde r)}.
\end{equation}
One can define a quantity, $\tilde \eta_{xy}$ on the $r=\tilde r$ hyper surface as
\begin{equation}
\label{shear at the hyper surface}
\tilde \eta_{xy}(\tilde r, \nu) \equiv -\frac{G_{\Psi}(\tilde r, \nu)}{i\nu},
\end{equation} 
for the future convenience.
Kubo formula for the shear viscosity cares the imaginary part of the retarded 
Green's function only. Therefore, the shear viscosity $\eta_{xy}$ is given by
\begin{equation}
\eta_{xy}\equiv Re[\tilde \eta_{xy}].
\end{equation}

The holographic $RG$ flow equation of the shear viscosity can be derived from the bulk equations of motion. 
Using Eq(\ref{Greens function at the hyper surface}) and 
Eq(\ref{shear at the hyper surface}), we switch $\Psi_\nu(r)$  in Eq(\ref{Psi-equation}) to $\tilde \eta_{xy}$, then we get
\begin{equation}
\label{xy-flow equation}
\partial_{\tilde r}\tilde \eta_{xy}=\frac{i\nu}{\sigma(\tilde r)N(\tilde r)} \left[ \frac{2\kappa^2_5}{\tilde r^3f^6(\tilde r)}\tilde\eta^2_{xy}
-\frac{\tilde r^3f^6(\tilde r)}{2\kappa^2_5}\right]+\alpha^2 \Delta(\tilde r),
\end{equation}
where
\begin{equation}
\Delta(r)\equiv\frac{rN(r)\sigma(r)f^4(r)}{i\kappa^2_5 \Psi_{\nu}(r) \nu}\left( \omega^{\prime}(r)\delta A_{y}^{1\prime}(r) -\frac{\omega(r)\phi^{2}(r)\delta A_{y}^{1}(r) }{N^{2}(r)\sigma^{2}(r)} +\frac{i\nu\omega(r)\phi(r)\delta A_{y}^{2}(r) }{N^{2}(r)\sigma^{2}(r)}\right).
\end{equation}
On the right hand side of Eq(\ref{xy-flow equation}), two terms in the square bracket will be the only terms 
remaining when the the black brane stays above its critical temperature, $T>T_c$(The background geometry 
becomes isotropic). 
In zero frequency limit, $\nu \rightarrow 0$, these terms vanish. Then, Eq(\ref{xy-flow equation}) becomes $\partial_{\tilde r}\tilde\eta_{xy}=0$ 
and 
the $RG$-flow is trivial. In this case, the shear viscosity becomes the universal value for any $r=\tilde r$ hyper surfaces.

However, below the critical temperature, the background geometry becomes anisotropic and the $\Delta(r)$ term in 
Eq(\ref{xy-flow equation}) appears, 
which comes from interactions with perturbative Yang-Mills fields. Therefore, it is manifest that in the zero frequency limit, 
the holographic $RG$ flow equation of the shear viscosity is not trivial as the usual cases \cite{Liu1,Zhou1}, 
in fact it flows from its $UV$ boundary value(\ref{ratio eta_yz and s}) to another one all the way to the $IR$ 
region.

The shear viscosities deeply in $IR$ region in the dual field theories correspond to those at the black brane horizon in the bulk gravity system.
In general ground, it is expected (e.g. see \cite{Anton1}) that the shear viscosities at the horizon in $SO(3)$ rotational symmetry broken direction 
will be modified from its universal value by 
the ratios of the metric components in the unbroken symmetry directions to those of the broken directions. In our case, these might be given  by
\begin{equation}
\label{shear-at-horizon}
\eta_{xy}(r=1)=\frac{1}{2\kappa^2_5}\frac{G_{yy}(r=1)}{G_{xx}(r=1)} {\rm\ \ and\ \ }\eta_{xz}(r=1)=\frac{1}{2\kappa^2_5}\frac{G_{zz}(r=1)}{G_{xx}(r=1)},
\end{equation}
where again $r=1$ is the location of the black brane horizon and $\frac{1}{2\kappa^2_5}$ is the universal value of the shear viscosity. In the following,
we will argue that this expectation is indeed right even in the case that the shear viscosity runs non-trivially.

We firstly investigate the near horizon limit of the equations of motion (\ref{Psi-equation}).
It turns out that the contributions from interaction term $\Delta(r)$ becomes relatively weak near horizon. 
Once we impose ingoing boundary condition at the horizon for the fields appearing in  Eq(\ref{Psi-equation}), 
Eq(\ref{a1-equation}) and Eq(\ref{a2-equation}) as 
$\Psi \sim \delta A^1_y \sim \delta A^2_y \sim (r-1)^{-i\nu \beta}$, where $\beta$ is a positive real number, 
the terms involving $\delta A^1_y$ and $\delta A^2_y$ in Eq(\ref{Psi-equation}) become less singular than the terms 
which are proportional to $\Psi_\nu$ and its derivatives. Then, dominant behavior in Eq(\ref{Psi-equation}) 
presents approximately near horizon limit of the equation in isotropic phase as
\begin{equation}
 0 \eqsim \Psi^{\prime\prime}_\nu(r)
+\frac{\Psi^{\prime}_\nu(r) }{r-1}
+\frac{\nu^2\Psi_\nu(r)}{N^{2}(r)\sigma^{2}(r)}.
\end{equation}
In fact, the solution of the gravitational field, $\Psi_\nu(r)$ has the following near horizon form in small frequency limit:
\begin{equation}
\label{near horizon Psi}
\Psi_{\nu}(r)=A_0(r,\nu)(r-1)^{-\frac{i\nu}{N_0 \sigma_0}}\sum_{n=0}^{\infty}\nu^n\Psi_n(r),
\end{equation}
where $ N_0  \equiv lim_{r \rightarrow 1}\frac{N(r)}{r-1}$, $ \sigma_0 \equiv \sigma(r=1)$, and 
$A_0(r,\nu)$ and $\Psi_n(r)$ 
are some regular functions in $r$ and their derivatives are regular too. 
Once we plug Eq(\ref{near horizon Psi}) into the definition of retarded Green's function(\ref{Greens function at the hyper surface}), it becomes
\begin{equation}
 G_{\Psi}(r,\nu)=\frac{r^3\sigma(r)N(r)f^6(r)}{2 \kappa^2_5}\partial_{r}\left[-\frac{i\nu}{N_0  \sigma_0} ln(r-1) + ln A_0(r,\nu)+ln \left( \sum_{n=0}^{\infty}\nu^n\Psi_n(r) \right)\right].
\end{equation}
The last two terms in the square bracket are regular functions in $r$ and $N(r)$ presents single zero at the horizon. 
Therefore, the first term in the square bracket is only surviving at the horizon, which gives
\begin{equation}
\label{Greens function at the horiZON}
G_{\Psi}(r=1,\nu)=-\frac{i\nu f^6(r=1)}{2\kappa^2_5} {\rm \ \ and \ \ }\eta_{xy}=\frac{1}{2\kappa^2_5}f^6(r=1).
\end{equation}  
The factor $f^6(r=1)$ in the Green's function and $\eta_{xy}$ is precisely the ratio, $\left.\frac{G_{yy}(r)}{G_{xx}(r)}\right|_{r=1}$ 
as expected in (\ref{shear-at-horizon}). 

One can derive the same result by using Eddington-Finkelstein coordinate in the gravity system. Requesting ingoing boundary condition is imposing regularity at the 
black brane horizon. Therefore, the any fields at the horizon depends on $t$ and $r$ only through their non-singular combinations. The ingoing null coordinate $v$ is
given by
\begin{equation}
dv=dt+\frac{dr}{N(r)\sigma(r)},
\end{equation}
and this implies 
\begin{equation}
-i\nu \Psi_\nu(r)=N(r)\sigma(r)\partial_r \Psi(r).
\end{equation}
Plugging this relation into (\ref{Greens function at the hyper surface}), one obtains the same Green's function with (\ref{Greens function at the horiZON}).


\subsection{$RG$-Flow Example of the Shear Viscosity from Analytic Solutions}
\label{The Exact $RG$-Flow Solution of the Shear Viscosity in the Perturbative Regim}
In this subsection, we provide an example of $RG$ flow from the analytic solutions to support the argument in 
Sec.\ref{General}.
The analytic solution of the bulk equations of motion is already obtained in \cite{Oh1}\footnote{See Appendix.C in \cite{Oh1}} and we
will use its result. 
Up to leading order corrections
in the double expansion of $\varepsilon \tilde D_{1}$ and $\alpha^2$, we obtained a radial flow of the retarded
Green's function by using Eq(\ref{Greens function at the hyper surface}) as a function of $r$:
\begin{eqnarray}
G_{\Psi}(r,\nu)&=&-\frac{i\nu}{2\kappa^2_5}
+ \frac{\alpha^2\varepsilon^2}{\kappa^2_5}\left[ \frac{59\tilde D^2_1 r^2(r^2-1)^2}{56(1+r^2)^4}
+\frac{\nu \tilde D_1}{5376(1+r^2)^4}\left( 10752 r^2(r^2-1)^2 
\frac{\tilde  A^{(1)}_{1,0}}{\tilde \Psi} \right.\right. \\ \nonumber
&+&\left.\left.i\tilde D_1(1193-2380r^2+726r^4-348r^6-87r^8+336r^2(r^2-1)^2 ln(1+\frac{1}{r^2}))\right)\right]\\ \nonumber 
&+&{\rm \ higher\ order \ in \ \nu,\ \varepsilon \ or\ \alpha^2},
\end{eqnarray}
where $\tilde \Psi = lim_{r\rightarrow \infty} \Psi_\nu(r)$, $\tilde A^{(1)}_{1,0}$ is 
an integration constant, 
which is hard to be determined analytically\footnote{See Eq(72),
Appendix.C in \cite{Oh1}}. 
So we keep $ \tilde A^{(1)}_{1,0}$ as unknown in this discussion.
We note that $\varepsilon^2 \alpha^2 = \frac{1260\pi(T-T_c)}{17\tilde D^2_1}$, so the temperature dependence  
of the retarded Green's function is encoded in this term.
Only imaginary part of the retarded Green's function contributes to the shear viscosity, so the first term in the square bracket 
does not contribute since it is purely real.
The shear viscosity on $r=\tilde r$ hyper surface is given by
\begin{eqnarray}
\eta_{xy}&=&\frac{1}{2\kappa^2_5}
-\frac{ \alpha^2\varepsilon^2\tilde D_1}{5376\kappa^2_5(1+\tilde r^2)^4}\left( 10752 \tilde r^2(\tilde r^2-1)^2 
 Im \left[\frac{\tilde A^{(1)}_{1,0}}{\tilde \Psi}\right] \right. \\ \nonumber
&+&\left.\tilde D_1(1193-2380r^2+726r^4-348r^6-87r^8+336r^2(r^2-1)^2 ln(1+\frac{1}{r^2}))\right)\\ \nonumber 
&+&{\rm \ higher\ order \ in \ \nu,\ \varepsilon \ or\ \alpha^2}.
\end{eqnarray}
Since the entropy density is constant in radial direction, $s=\frac{2\pi}{\kappa^2_5}$, the ratio of the shear viscosity and
entropy density runs non-trivially too.
As $\tilde r \rightarrow \infty$, 
the Green's function and the shear viscosity become 
\begin{equation}
G_{\Psi}(r=\infty,\nu)=-\frac{i\nu}{2\kappa^2_5}\left( 1+ \frac{29}{896}\varepsilon^2\alpha^2 \tilde D^2_1\right) 
{\rm \ and \ } \eta_{xy}(r=\infty)=\frac{1}{2\kappa^2_5}\left( 1+ \frac{29}{896}\varepsilon^2\alpha^2 \tilde D^2_1\right),
\end{equation}
up to leading order correction. 
At the black brane horizon $\tilde r=1$, the retarded Green's function and the shear viscosity flows into the values as
\begin{equation}
G_{\Psi}(\tilde r=1,\nu)=-\frac{i\nu}{2\kappa^2_5}\left( 1+\frac{1}{48}\varepsilon^2\alpha^2 \tilde D^2_1 \right) 
{\rm \ and \  }\eta_{xy}(\tilde r=1)=\frac{1}{2\kappa^2_5}\left( 1+\frac{1}{48}\varepsilon^2\alpha^2 \tilde D^2_1 \right).
\end{equation}
The factor causing deviations from the universal value of the shear viscosity at the horizon is 
\begin{equation}
1+\frac{1}{48}\varepsilon^2\alpha^2 \tilde D^2_1=f^6(r=1)=\frac{G_{yy}(r=1)}{G_{xx}(r=1)},
\end{equation}
in this perturbative regime, which is exactly as expected in (\ref{shear-at-horizon}).

\section*{Acknowledgements}
I would like thank to Shesansu Shekar Pal, Sayantani Bhattacharyya, Ashoke Sen, Dileep Jatkar for the discussion. This work is mostly developed in
CQUeST(Center for Qnatum Spacetime) in Sogang University. I thank to everyone in CQUeST for hospitality, especially 
to Bum-Hun Lee for invitation. I have also developed this idea by some private discussions 
in National String Meeting(NSM)2011 in New Delhi. I would like thank to the staffs and the organizers. I also thank to Patrick Kerner and the referee 
in {\it JHEP} who commented on my previous submission. They individually inform me that
the conclusion in the previous draft is partially incorrect and lets me have chance to consider that again. Finally, I would like thank
to my $\mathcal{W.J.}$

This work is partially supported by 11-R$\&$D-HRI-5.02-0304 through Harish-Chandra Research
Institute in India.

\end{document}